\documentclass[twocolumn,twocolappendix]{aastex63}

\usepackage{amsmath}
\usepackage{soul}
\usepackage{ulem}
\usepackage{multirow}
\definecolor{dgreen}{RGB}{26,148,49}
\definecolor{forest}{RGB}{3, 148, 49}

\usepackage{lineno}

\shorttitle{Dipole Tension and Clustering}
\shortauthors{Arefeh Daei Rasouli, Haniyeh Sadat Tadayyoni, et al.}

\begin{document}

\title{Addressing Dipole Tension via Clustering in $\Lambda$CDM and beyond}

\author{Arefeh Daei Rasouli\textsuperscript{†}}

\affiliation{Department of Physics, Sharif University of Technology, Tehran 11155-9161, Iran}

\author{Haniyeh S. Tadayyoni\textsuperscript{†}}

\affiliation{Department of Physics, Sharif University of Technology, Tehran 11155-9161, Iran}

\footnotetext[1]{\textsuperscript{†} These authors contributed equally to this work.}

\author[0000-0001-6131-4167]{Shant Baghram}
\affiliation{Department of Physics, Sharif University of Technology, Tehran 11155-9161, Iran}
\address{Research Center for High Energy Physics,
Department of Physics,
Sharif University of Technology,
Tehran 11155-9161, Iran}

\author[0000-0002-7084-5725]{Sohrab Rahvar}
\affiliation{Department of Physics, Sharif University of Technology, Tehran 11155-9161, Iran}
\address{Research Center for High Energy Physics,
Department of Physics,
Sharif University of Technology,
Tehran 11155-9161, Iran}

\begin{abstract}
The dipole in the angular distribution of the cosmic microwave background (CMB) is attributed to the Doppler effect and our motion relative to the CMB rest frame. It is expected that observations of large-scale structures (LSSs) would also exhibit a related kinematic dipole. However, numerous studies of the LSS dipole have shown significant discrepancies with predictions based on the CMB. In this work, we investigate how considering the clustering dipole affects the LSSs distribution dipole using the National Radio Astronomy Observatory (NRAO) Very Large Array (VLA) Sky Survey (NVSS) and the Wide-field Infrared Survey Explorer (WISE), and examine the nonlinear regime to calculate the correlation between the clustering and the kinematic dipole. Our results show up to $\lesssim28\%$ enhancement in the clustering dipole amplitude compared to previous studies, with increases of up to $\lesssim 22\%$ in $\Lambda$CDM and $\lesssim 28\%$ in modified gravity scenarios. 
Additionally, we explore a model in which the distribution of matter on LSS could be intrinsically anisotropic by a long-mode modulation. Using the remnant discrepancy between the observed and predicted dipole, we derive an upper limit for the amplitude of intrinsic dimensionless anisotropy $\lesssim 0.22$. Furthermore, we investigate these results within the framework of the $f(R)$  modified gravity model. We conclude that nonlinear clustering and local structure correlations partially alleviate the dipole tension within $\Lambda$CDM, yet, this anomaly remains a challenge. Two alternative models are in the direction of relaxing the tension. However, further investigation and more accurate data are needed to support a stronger argument.

\end{abstract}


\newpage
 \section{Introduction} \label{sec:intro}
The standard model of cosmology, based on the assumption of a cosmological constant $\Lambda$ as the cause of the accelerated Universe, and cold dark matter (CDM) as the main component of the Universe's structures, is known as $\Lambda$CDM. This model, $\Lambda$CDM, is based on the cosmological principle that the Universe is homogeneous and isotropic on large scales. In accordance with one uses the Friedmann–Lemaître–Robertson–Walker (FLRW) metric, initially motivated by simplicity and as empirical evidence of the cosmological principle. This has since become a foundational pillar of modern cosmology \citep{2020coce.book.....P}.
		While direct tests of homogeneity, such as counts-in-cells analyses, generally support the cosmological principle on very large scales, the persistence of our local bulk flow on large scales remains an area of active investigation. The assumption of isotropy is supported by observations of cosmic microwave background (CMB) temperature fluctuations of amplitude $\Delta T / T \sim 10^{-5}$ \citep{Akrami2020b}. 
According to the standard picture of structure formation through gravitational instability, the large-scale structure matter distribution originates from the nearly Gaussian density fluctuations created in the early Universe, which represent the higher multipoles of CMB we observe today \citep{2020A&A...641A..10P, 2020A&A...641A...9P}. This suggests that the large-scale distribution of matter in the linear regime and the CMB should have the same rest frame. 

The dipole anisotropy of the CMB observed from the solar system barycentric frame, however, shows the amplitude of $\sim 10^{-3}$. This dipole is commonly interpreted as a result of the Solar System's motion relative to the CMB rest frame, with a velocity of $369.82 \pm 0.11$ km/s directed toward Galactic coordinates $(l, b) = (264.021^\circ \pm 0.011^\circ, 48.253^\circ \pm 0.005^\circ)$\citep{2020AA...641A...1P}. A model-independent approach to validate this kinematic hypothesis is to determine the dipole moment in the angular distribution of large-scale structures.
\cite{Ellis1984} proposed this consistency test, providing a quantitative prediction: for a population of sources with spectral index $\alpha$ and number count index $x$, a moving observer should measure a kinematic dipole amplitude. If we move with velocity $v \ll c$ concerning the isotropic distribution of distant radio galaxies, a dipole anisotropy will be observed in the distribution of sources. This dipole effect arises because relativistic aberration causes a concentration of sources in the direction of our velocity \citep{Bradly}, and the relativistic Doppler boosting modifies the observed frequency of sources.

The release of the National Radio Astronomy Observatory (NRAO) Very Large Array (VLA) Sky Survey (NVSS; \citealt{1998AJ....115.1693C}) catalog made it possible to obtain the first accurate measurements of the matter dipole. \cite{2002MNRAS.329L..37B} carried out the initial study using this dataset. Later, \cite{2011ApJ...742L..23S}  confirmed that the dipole direction is consistent with that of the CMB, but also found that its amplitude was larger than expected. This discrepancy was further studied by \cite{2012MNRAS.427.1994G} with independent methods, and by other groups using additional radio surveys such as the Westerbork Northern Sky Survey (WENSS) \citep{Siewert2021}, the Sydney University Molonglo Sky Survey (SUMMS) \citep{2017MNRAS.471.1045C}, and the TIFR Giant Metrewave Radio Telescope (TGSS) Sky Survey \cite{Bengaly2018}. Together, these works showed that the anomaly is persistent across different radio datasets.
		Independent support for this finding came from infrared data. \cite{2021ApJ...908L..51S} analyzed active galactic nuclei (AGNs) detected by the Wide-field Infrared Survey Explorer (WISE; \citealt{2010AJ....140.1868W}) mission and reported a dipole amplitude significantly larger than the kinematic prediction. Their later study \citep{Secrest2022} confirmed this result, providing strong multi-wavelength evidence that the anomaly cannot simply be explained by systematics specific to radio observations.
        
This tension between the CMB and large-scale structure rest frame remains a challenge to the $\Lambda$CDM model. Recent reviews, such as \cite{Perivolaropoulos2022} and \cite{Valentino2025}, provide a comprehensive overview of the challenges facing the $\Lambda$CDM model, examining persistent tensions across multiple cosmological probes and discussing potential systematic effects and theoretical extensions.

In this work, to reassess this discrepancy, we assume that the kinematic dipole \textemdash  a dipole results from the observer’s motion relative to the cosmic rest frame \textemdash of the galaxy distribution is consistent with the CMB dipole. We then analyze the shot noise and clustering dipole \textemdash a dipole results from uneven distribution of matter and galaxies in the Universe \textemdash using the clustering redshift method and cross-matching technique, following the idea introduced by \cite{Cheng2024}. For this purpose, a probability distribution for the NVSS dipole in the standard cosmological model and within the linear regime was derived, including the kinematic, clustering, and shot noise components, which was found to be consistent within approximately $2\sigma$ accuracy \citep{Cheng2024}.

In this research, to improve the results and to examine the effects of the nonlinear regime compared to the model obtained by \cite{Cheng2024}, we analyze the impact of the nonlinear regime on the clustering dipole within the standard cosmological model. Additionally, we consider two types of paradigms as alternatives to $\Lambda$CDM: one is long-mode modulation in the early Universe \citep{1991PhRvD..44.3737T}, and the other is the $f(R)$ model as a late-time modification \citep{2007PhRvD..75d4024B,2009PhRvD..80f4003B,2010LRR....13....3D,2010RvMP...82..451S,2012PhR...513....1C}, which has been investigated in both linear and nonlinear regimes.

Throughout this paper we use the $\Lambda$CDM cosmological parameters from the \textit{Planck} TT,TE,EE+lowE+lensing results \citep{2020AA...641A...6P}: we set CDM density $\Omega_c h^2=0.120$, baryon density $\Omega_b h^2=0.0224$, Hubble parameter $H_0=100h\, \text{km\,s}^{-1}\text{Mpc}^{-1}$, with $h=0.674$, primordial comoving curvature perturbation power spectrum amplitude $A_s=2.1 \times 10^{-9}$, defined at the pivot scale $k_p=0.05\,\text{Mpc}^{-1}$, and we expect our results to be only very weakly dependent on these parameters.

The structure of this paper is organized as follows: In Sections \ref{sec:TB}, we introduce the components of the dipole observed in the distribution of galaxy sources, along with the methods we use to measure them. In Section~\ref{sec:Results}, we present the amplitude derived for the clustering term using both the NVSS survey and CatWISE catalog of quasars, along with the effect of including the correlation between the kinematic and clustering terms, all while considering nonlinear corrections to the matter power spectrum. Finally, in section~\ref{sec:Beyond}, we discuss two approaches to address the remaining tension: a simple model that allows for intrinsic anisotropies, and a beyond $\Lambda$CDM scenario. We discuss the theories we applied to consider nonlinear corrections in Appendix \ref{sec:Nonlinear}.
\section{Theoretical Background}\label{sec:TB}
Under the assumption that both the LSS and the CMB have the same rest frame, the dipole signal observed in the LSS can be interpreted as a combination of distinct contributing factors.
The total observed dipole, \(\mathbf{d}\), quantifies the anisotropy in the angular distribution of sources in the sky and represents the first-order term in the spherical harmonic expansion of the source distribution. This dipole can be decomposed into three primary components: the kinematic \(\mathbf{d}_{\text{kin}}\), the clustering \(\mathbf{d}_{\text{clus}}\) and the shot-noise dipole \(\mathbf{d}_{\text{SN}}\):
\begin{equation}
    \label{eq:total_dipole}
    \mathbf{d} = \mathbf{d}_{\text{kin}} + \mathbf{d}_{\text{clus}} + \mathbf{d}_{\text{SN}}.
\end{equation}
The kinematic dipole results from the observer’s motion relative to the cosmic rest frame, typically described by the Doppler effect and the aberration of light. The clustering dipole, on the other hand, arises from the uneven distribution of matter and galaxies in the Universe. This anisotropy is not a result of the observer’s motion but is intrinsic to the large-scale structure, as gravitational potential wells associated with overdense regions influence the distribution of galaxies and other tracers of matter. In essence, the clustering dipole reflects the alignment of observed structures along certain directions due to these density fluctuations, distinct from the motion-induced anisotropies of the kinematic dipole. Finally, the shot-noise dipole arises from the discrete sampling of sources, which introduces statistical fluctuations in the observed signal due to finite sampling. However, these fluctuations should, on average, be isotropic when the source distribution is appropriately normalized across all directions. Any residual dipole signal attributed to shot noise would therefore reflect deviations from perfect normalization or incomplete sampling. In the following, we will explore each of these dipole components in detail.
\subsection{Kinematic Dipole}\label{ssec:Kinematic}
The kinematic dipole arises from the observer's motion relative to the CMB rest frame, introducing a dipolar modulation in the observed number counts of sources due to relativistic Doppler and aberration effects. When an observer moves with velocity \( \mathbf{v_k}  \) relative to the CMB rest frame, the isotropic distribution of sources appears anisotropic: a higher observed count is noted in the direction of motion, while a lower count appears in the opposite direction. This asymmetry is further influenced by aberration, which shifts the apparent positions of sources, altering their observed distribution.
All sources are assumed to follow a power-law spectral profile, where the flux density  \( S(\nu) \)  depends on frequency as  \( S(\nu)\propto \nu^{-\alpha} \), with  \( \alpha \)  as the spectral index. For galaxies, this spectral behaviour is commonly observed, with values of  \( \alpha \)  typically ranging between 0.5 and 1.0 for radio emissions \citep{Condon1992}. The average number of sources above some limited flux density \( S \) also follows a power-law distribution given by:
\begin{equation}
	\frac{dN}{d\Omega}(> S) = k S^{-x},
	\label{eq:fluxpowerlaw}
\end{equation}
where \( x \) represents the slope of the source count distribution. This distribution assumes \( S \) represents the total flux integrated over the relevant frequency range. Observationally, this requires corrections for the source's spectral energy distribution (SED), which determines energy emission across wavelengths, and redshift effects, which shift emitted light to longer wavelengths, altering the observed flux. These considerations are essential to ensure the equation aligns with observational data and accurately reflects the statistical distribution of sources.

In the observer’s frame, these Doppler and aberration effects together generate a measurable dipole anisotropy in the observed source counts. The integral source counts above a flux threshold \( S \), when observed from a frame moving at velocity \( v_k \) relative to the cosmic rest frame, are given by
\begin{equation}
	\label{eq:dN/dOmega}
	\left(\frac{dN}{d\Omega}\right)_{\text{obs}} = \left(\frac{dN}{d\Omega}\right)_{\text{rest}} \delta^{2 + x(1+\alpha)},   
\end{equation}
where \( \delta = [1 + (v_k/c)\cos\theta] / \sqrt{1 - v_k^2/c^2} \) is the relativistic Doppler factor, and \( x \) and \( \alpha \) are the slope of the source count distribution and the spectral index, respectively. The factor \( 2 + x(1 + \alpha) \) arises from the combined contributions of Doppler modulation and aberration.
From this modulation of the number counts, the kinematic dipole has an amplitude
\begin{equation}
	d_{\text{kin}} = \beta[2 + x(1+\alpha)], 
	\label{eq:d_k}
\end{equation}
where \( \beta = v_k/c \). This amplitude quantifies the first-order effects of the observer’s velocity relative to the cosmic rest frame, encompassing both the alteration in flux and the redistribution of source positions across the sky. By appropriate normalization, the expression ensures consistency with the isotropy of source counts in their rest frame.
This framework is consistent with the principles of the \cite{Ellis1984} test, which examines whether the kinematic dipole resulting from the observer's motion relative to the rest frame of sources aligns with the dipole anisotropy determined from the CMB.

Some previous studies have suggested possible corrections to the \cite{Ellis1984} kinematic dipole prediction, {{such as redshift-dependent or spatial variations in the spectral index \(\alpha(z)\) \citep{Dalang2022} or deviations from a strict power-law in the source counts \citep{Tiwari2015, Rubart2013}. However, \cite{2024MNRAS.535L..49V} demonstrated that the \cite{Ellis1984} test is robust to such effects and the Equation~(\ref{eq:d_k}) remains valid when the effective spectral index \(\alpha\) and source count slope \(x\) are measured at the applied flux limit. This is because the kinematic dipole arises from sources crossing the survey’s flux limit in a boosted frame, meaning that variations of \(\alpha(z)\) or deviations of the integral counts far from the flux limit do not affect the predicted kinematic amplitude.}}

\subsection{Clustering Dipole}\label{ssec:clustering}
An anisotropic distribution around an observer will produce a nonnegligible dipole called the clustering dipole. The amplitude of this dipole,
$d_{\text{clus}}$ can be computed given the angular distribution of the galaxies:
$d_{\text{clus}}=\sqrt{9 C_1 / 4\pi}$, where the clustering power spectrum $C_{\ell}$ is modeled as \citep{Tegmark_2002}:
\begin{equation}
	\label{eq:cl1}
	C_{\ell} =\frac{2}{\pi}  \int_0^{\infty} dk \ k^2 P(k; z=0)  \left[\int_0^{\infty} d\chi \ j_{\ell}(k\chi) F(\chi) \right]^2.
\end{equation}
Here $j_{\ell}$ is the spherical Bessel function, and the radial selection function, $F(\chi)$, is given by:
\begin{equation}
	\label{eq:F(chi)}
	F(\chi) \propto D(z) b(z) \frac{dN}{dz}(\chi),
\end{equation}
where $D(z)$ represents the linear growth function, normalized to unity at present. Additionally, $b(z)$ is the clustering bias, and $\frac{dN}{dz}$ corresponds to the surface density of sources per unit redshift $z(\chi)$. Since we are interested in studying the effect of small scales on clustering, we include both the linear and nonlinear matter power spectrum $P(k)$ in our calculations. In Equation~(\ref{eq:F(chi)}), the redshift dependency of the power spectrum is considered in $F(\chi)$ through the growth function parameter and the bias parameter. 
The light from sources at cosmological distances that we observe corresponds to the past light cone; consequently, the angular clustering power spectrum evolves with the unequal-time matter power spectrum as we consider the time evolution of the density field. The resulting clustering power spectrum is:
\begin{align}
	\label{eq:uneq}
	C_{\ell} &= \frac{2}{\pi} \int dk k^2  \int  d\chi F(\chi) j_{\ell}(k\chi) \notag \\
	& \int  d\chi^{\prime} F(\chi^{\prime}) j_{\ell}(k\chi^{\prime}) P(k; \eta(\chi), \eta(\chi^{\prime})),
\end{align}
where $\eta$ represents the conformal time, $\eta(\chi)=\eta_0 - \chi(z)$. The clustering dipole values utilizing the unequal-time matter power spectrum have been calculated using Equation~(\ref{eq:uneq}).
We consider the mean geometric method \citep{Castro2005, Kitching2017} for calculating $P(k; \eta(\chi), \eta(\chi^{\prime}))$:
\begin{equation}
	\label{eq:Appr}
	P(k; z_1, z_2) \simeq  [P(k; z_1) P(k; z_2)]^{\frac{1}{2}}.
\end{equation}
This computes the correlation of the density field within each redshift interval and simplifies the numerical integration of the angular clustering power spectrum. This approximation is accurate at large and even mildly nonlinear scales, where the linear matter power spectrum dominates over nonlinear corrections.
The values obtained for the clustering dipole using the geometric mean method are represented in Tables \ref{tab:clustering_dipole_Approx_unequaltime_one_loop} and \ref{tab:clustering_dipole_CCL} for $\Lambda$CDM, and in Tables \ref{tab:clustering_dipole_CCL_MG1} and \ref{tab:clustering_dipole_CCL_MG2} for $f(R)$ gravity model.

To find the clustering values, we use the redshift distribution and bias, $b(z)$, of both the NVSS radio survey and the CatWISE catalog of quasars.
For NVSS catalog, we employ cross-correlation techniques
\citep{Newman2008, Schmidt2013, Mnard2013, Chiang2019}; the redshift distribution is estimated through spatial cross-correlation analyses between two sets of objects, one with known accurate redshifts (the spectroscopic map) and the other with unknown redshifts (the photometric map). In this study, we use the derived $b(z) dN/dz$ of the NVSS catalog by
\cite{Cheng2024} and model the fiducial redshift distribution by fitting a step function to it. As explained in \cite{Cheng2024}, the fiducial model we adopt multiplies their minimum model redshift distribution by a factor of three to better match tomographic bias estimates for low redshift. This correction compensates for the clustering bias neglected in the minimum model and potential incompleteness in the cross-matched sample, leading to a slightly higher clustering amplitude.
\cite{Cheng2024} employed the \textit{Tomographer} package\footnote{\url{http://tomographer.org/}} to conduct the clustering redshift calculations.
This was done by cross-correlating the NVSS source catalog with compiled spectroscopic samples of galaxies and quasars from the Sloan Digital Sky Survey \citep{Strauss2002, Schneider2010, Paris2018}. Since \textit{Tomographer} only included redshifts greater than 0.06, they cross-matched with the 2MASS Redshift Survey \citep{Huchra2012} and the local
radio sources to establish a lower limit for the redshift intervals. 

There are known observational systematics related to the NVSS catalog that can bias large-scale dipole analyses and clustering measurements. These include declination-dependent variations in sensitivity and rms noise, which produce surface-density fluctuations of sources across the sky, flux calibration and direction-dependent effects of the calibration, incompleteness and survey masking (in particular near the Galactic plane), estimator bias and shot-noise effects for sparse and final samples, and resolution or blending of extended sources that can alter counts and fluxes. Addressing these systematics is required to properly evaluate NVSS dipole and clustering measurements. Evidence and discussion of these effects can be found in more detail in the survey description, \cite{1998AJ....115.1693C}, and follow-up studies \citep{2016A&A...591A.135C, Rubart2013, 2016JCAP...03..062T, Bengaly2018}.

For CatWISE catalog, we follow the methodology of \cite{2021ApJ...908L..51S} to construct our quasar sample, based on the CatWISE2020 data release \citep{2020ApJS..247...69E}. After applying the standard selection criteria, we cross-match with SDSS Stripe 82 to obtain a spectroscopic subsample of $7734$ sources. We fit an analytical function to their redshift distribution to derive $dN/dz$, calculating Poisson errors for the binned data. The bias parameter is set to $b = 1.0$, adopted directly from \cite{2021ApJ...908L..51S}.
{{The CatWISE catalog is also affected by observational systematics, including nonuniform sky coverage that causes position-dependent variations in depth and completeness, photometric uncertainties, and contamination from stars and artifacts, which can induce sky-dependent fluctuations in source counts and thereby affect the clustering values. 
Further details regarding these systematics can be found in \cite{2020ApJS..247...69E}.
}}

\subsection{Shot-noise Dipole}\label{ssec:Shot-noise}
The shot-noise dipole component arises from the discrete sampling of point sources, which introduces randomness due to their finite number. For a large number of sources (\( N \rightarrow \infty\)), the underlying field could be accurately reconstructed, but with limited sources, shot-noise affects the observed dipole. Assuming full-sky coverage, the field of \( N \) point sources with random and isotropic positions, denoted as \(\hat{n}_i\), can be described using Dirac delta functions centered at each source location. This discrete nature of the sources introduces shot-noise in the angular power spectrum, which becomes significant for a finite number of sources.
For the shot-noise contribution, the angular power spectrum \( C_l^{\rm{SN}}\) is constant for \(l>0\) and is given by:
\begin{equation}
\label{eq:Cl_SN}
    C_l^{\rm{SN}}=\frac{1}{\bar{N}},
\end{equation}
Where \(\bar{N}=N/4\pi\) is the mean source density. In the limit of \( N \rightarrow \infty\), shot-noise vanishes, but with finite \(N\), the shot-noise dipole amplitude \(d_{\rm{SN}}\) for \(l=1\) can be estimated by:
\begin{equation}
    \label{eq:d_SN}
    d_{\rm{SN}}=\sqrt{\frac{9C_1^{\rm{SN}}}{4\pi}}.
\end{equation}
To compute the shot-noise angular power spectrum \( C_l^{\rm{SN}}\), we follow the methodology from \cite{Blake2004}, which incorporates the impact of multicomponent sources, such as double radio galaxies, on the power spectrum. \cite{Cheng2024} report the shot-noise dipole amplitude for NVSS to be \(d_\text{SN}=0.46\times 10^{-2}\).
\section{Results: Dipole in the Linear and Nonlinear Regimes of \texorpdfstring{$\Lambda$}{Lambda}CDM }\label{sec:Results}
\nolinenumbers
\vspace{-2.0em} 
\begin{table*}[t]
\centering
\begin{tabular}{|c|c|c|c|c|c|c|c|}
    \hline
    {Galaxy Sample} & {Redshift} 
    & \multicolumn{2}{c|}{$0.01 < z < 1$} 
    & \multicolumn{2}{c|}{$0.01 < z < 3$} 
    & \multicolumn{2}{c|}{$0.0025 < z < 3.14$} \\
    \hline
    \multirow{2}{*}{\shortstack{NVSS\\ \scriptsize(Radio Galaxies)}} 
    & $P(k)$ & Linear & Nonlinear & Linear & Nonlinear & Linear & Nonlinear \\
    \cline{2-8}
    & $d_{\text{clus}}(\times 10^{-2})$ 
    & $0.381^{+0.008}_{-0.005}$ & $0.382^{+0.008}_{-0.005}$ 
    & $0.386 \pm 0.007$ & $0.39 \pm 0.01$ 
    & $0.58 \pm 0.02$ & $0.62 \pm 0.04$ \\
    \hline
    \multirow{3}{*}{\shortstack{CatWISE\\ \scriptsize(Qusars)}} 
      & {Redshift} & \multicolumn{6}{c|}{$0 \lesssim z < 3$} \\
    \cline{2-8}
      & $P(k)$ & \multicolumn{3}{c|}{Linear} & \multicolumn{3}{c|}{Nonlinear} \\
    \cline{2-8}
      & $d_{\text{clus}}(\times 10^{-2})$ 
      & \multicolumn{3}{c|}{$0.024 \pm 0.003$} 
      & \multicolumn{3}{c|}{$0.025 \pm 0.003$} \\
    \hline
\end{tabular}
\caption{Clustering dipole values using the geometric mean approximation in the angular power spectrum (the 1-loop model).}
\label{tab:clustering_dipole_Approx_unequaltime_one_loop}
\end{table*}
\begin{center}
\begin{table*}[t]
\begin{tabular}{|c|c|c|c|c|c|}
    \hline
    {Galaxy Sample} & {Redshift} 
    & \multicolumn{2}{c|}{$0.01 < z < 3$} 
    & \multicolumn{2}{c|}{$0.0025 < z < 3.14$} \\
    \hline
    \multirow{2}{*}{\shortstack{NVSS\\ \scriptsize(Radio Galaxies)}} 
    & $P(k)$ & Linear & Nonlinear & Linear & Nonlinear \\
    \cline{2-6}
    & $d_{\text{clus}}(\times 10^{-2})$ 
    & $0.41 \pm 0.01$ & $0.42 \pm 0.01$ 
    & $0.62 \pm 0.02$ & $0.65 \pm 0.04$ \\
    \hline
    \multirow{3}{*}{\shortstack{CatWISE\\ \scriptsize(Quasars)}} 
      & {Redshift} & \multicolumn{4}{c|}{$0 \lesssim z < 3$} \\
    \cline{2-6}
      & $P(k)$ & \multicolumn{2}{c|}{Linear} & \multicolumn{2}{c|}{Nonlinear} \\
    \cline{2-6}
      & $d_{\text{clus}}(\times 10^{-2})$ 
      & \multicolumn{2}{c|}{$0.027 \pm 0.003$} 
      & \multicolumn{2}{c|}{$0.028 \pm 0.003$} \\
    \hline
\end{tabular}
\caption{Clustering dipole values using the geometric approximation in the angular power spectrum (the Halofit model).}
\label{tab:clustering_dipole_CCL}
\end{table*}
\end{center}
\hspace*{\parindent}In order to derive the total amplitude of a matter distribution catalog like NVSS, we assume that the kinematic term of the dipole matches the dipole anisotropy of the CMB in both magnitude and direction. Based on measurements by \cite{2020AA...641A...1P}, the CMB dipole corresponds to $\beta = (1.23357 \pm 0.00036)\times 10^{-3}$. Therefore, the velocity of the solar system with respect to the CMB rest frame is  $369.82 \pm 0.11\ \text{km}\ \text{s}^{-1}$. This estimate can be translated to an amplitude of $0.471 \times 10^{-2}$ \citep{Cheng2024} using Equation~(\ref{eq:d_k}). 
This kinematic dipole is derived by employing $\alpha = 0.75 \pm 0.25$ and using the ``mask d'' value of $\bar{x}=1.04$ from \cite{Siewert2021}, where the flux density threshold of NVSS is also 15 mJy. 

In addition to the kinematic dipole, we introduced the formalism for computing the amplitude of the clustering dipole as detailed in \ref{ssec:clustering}, the results of which are shown in Tables \ref{tab:clustering_dipole_Approx_unequaltime_one_loop} and \ref{tab:clustering_dipole_CCL} for $\Lambda$CDM and Tables \ref{tab:clustering_dipole_CCL_MG1} and \ref{tab:clustering_dipole_CCL_MG2} show the corresponding values for the Modified Gravity (\(f(R)\) Hu-Sawicki model) that is considered in section~\ref{ssec:ModifiedGravity}.
To compute the clustering values, we use the matter power spectrum in both linear and nonlinear regimes. For the nonlinear part, we employ two approaches: Standard Perturbation Theory (SPT) and the Halo Model, which are described in more detail in Appendix \ref{sec:Nonlinear}.
We also have the amplitude of the shot-noise term, $d_\text{SN}=0.46\times 10^{-2}$, as mentioned in \ref{ssec:Shot-noise}. Using these terms, we derive the probability distribution function for the total NVSS dipole and its amplitude in this section.
In our NVSS analysis, we used the redshift distribution of sources between 0.0025 and 3.14 to study the effect of including low-redshift sources. In Table~\ref{tab:clustering_dipole_Approx_unequaltime_one_loop}, we showed that by including sources within the redshift range $1<z<3$, the amplitude of the NVSS clustering dipole increases by only $\sim 1$--$2\%$ in the linear and nonlinear regimes, respectively. 
			Including NVSS sources with redshifts above 3, up to 3.14, was only done to capture the full fiducial redshift distribution. As expected from theory, their contribution is minimal, and even the nonlinear correction does not significantly affect the clustering dipole amplitude. It is a reasonable result as, by moving to higher redshifts, the scale of non-linearities moves to smaller scales.
\subsection{Probability Distribution for the Dipole Amplitude}\label{ssec:pdf}
We consider the total dipole of NVSS sources to consist of three main components: the kinematic, clustering and shot-noise dipole, where the last two terms are referred to as the source-based dipole $\mathbf{d}_S$. We then assume that each of these components has a Gaussian distribution and use the chain rule to express the resulting probability:
\begin{equation}
    P(\mathbf{d}) = \int d^3 \mathbf{d}_{\text{kin}} \, \int d^3 \mathbf{d}_{\text{S}} \, P(\mathbf{d}|\mathbf{d}_{\text{kin}}, \mathbf{d}_{\text{S}}) P(\mathbf{d}_{\text{kin}}, \mathbf{d}_{\text{S}}).
\end{equation}
Here, the Gaussian distribution of the kinematic term is modeled with a known mean amplitude $d_{\text{kin}}$ (CMB dipole), while the source-based terms follow a zero-mean distribution. Given the uncertainty $\sigma_\text{S}$ produced by the clustering of sources and shot-noise, we have:
\begin{align}
    \label{eq:pdf}
    P(\mathbf{d}|\sigma_\text{S}) &= \int d^3 \mathbf{d}_{\text{kin}} \, P(\mathbf{d}_{\text{kin}}) \notag \\
    & \cdot \int d^3 \mathbf{d}_{\text{S}} \, P(\mathbf{d} | \mathbf{d}_{\text{kin}}, \mathbf{d}_{\text{S}}) P(\mathbf{d}_{\text{S}} | \mathbf{d}_{\text{kin}}, \sigma_{\text{S}}).
\end{align}
According to the definition of the amplitude of NVSS dipole, $P(\mathbf{d} | \mathbf{d}_{\text{kin}}, \mathbf{d}_{\text{S}}) = \delta_D(\mathbf{d}-(\mathbf{d}_{\text{kin}}+\mathbf{d}_{\text{S}}))$.
The uncertainty of the source-based dipole can be computed as:
\begin{equation}\label{eq:sigma_S}
\sigma^2_{\text{S}} \approx \frac{1}{f_{\text{sky}} }\frac{3}{4\pi}C_1^{\text{S}},
\end{equation}
where $f_{\text{sky}}$ is the observed fraction of the sky. For the NVSS, after applying the Galactic plane mask and the bright source mask used in \cite{Cheng2024}, we adopt $f_{\text{sky}}=0.63$. Additionally, $C_1^{\text{S}}$ denotes the sum of the angular power spectrum of clustering and shot-noise for $\ell=1$ modes. 
This formula is exact when considering the full-sky limit.
Finally, the probability distribution function of source-based uncertainty can be represented as:
\begin{equation}
\label{eq:p(sigma_S)}
P(\sigma_{\text{S}}) = \frac{\partial d_{\text{clus}}}{\partial \sigma_\text{S}}P(d_{\text{clus}})=\sqrt{\frac{4\pi}{C_1}} \sigma_{\text{S}} P(d_{\text{clus}}).
\end{equation}
Marginalizing over $P(\sigma_{\text{S}})$ is the final step in deriving the probability distribution function of the total NVSS dipole amplitude.
\subsection{Correlation of the Kinematic and Clustering Dipoles} \label{ssec:correlation}
The gravitational interactions with the inhomogeneous distribution of local sources influence our motion relative to the CMB's rest frame, leading to a correlation between the kinematic and clustering dipoles. To account for the effect of this correlation, we use the model presented by \cite{Dam2023} (originally applied in quasar dipole analysis), which was later adopted by \cite{Cheng2024} for the probability distribution of the total NVSS dipole. 
Assuming that the kinematic term of the NVSS dipole is the same as the CMB dipole and directed along the $\hat{z}$, we can rewrite the probability distribution function considering the correlation as follows:
\begin{align}
    \label{eq:pdfc}
    P(\mathbf{d}|\sigma_S) = & \frac{1}{2\pi \tilde{\sigma}_L^2 \sqrt{2\pi(\tilde{\sigma}_S^2 + \sigma_{\text{kin}}^2)}}\\
    & \times \exp{(-\frac{d_x^2 + d_y^2}{2\tilde{\sigma}_L^2})} \exp{\left(-\frac{(d_z - \tilde{d}_{\text{kin}})^2}{2(\tilde{\sigma}_S^2 + \sigma_{\text{kin}}^2)}\right)} \nonumber,
\end{align}
where $d_{x,y,z}$ are components of the total dipole at each $\hat{x}, \hat{y}, \hat{z}$ dirction. The $\sigma_{\text{kin}}^2$ is the variance of the kinematic term, and $\tilde{\sigma}^2_S \equiv (1-r_c^2)\sigma^2_S$ represents the modified deviation of source-based terms, where $r_c$ is the correlation coefficient that will be introduced in the following. By considering the correlation between the clustering and kinematic dipoles, the amplitude of the kinematic dipole is displaced as $\tilde{d}_{\text{kin}}\equiv d_{\text{kin}} + d_c$, 
which can increase the amplitude of the total dipole and reduce the dipole tension.
\begin{figure}[ht!]
	\includegraphics[width=0.45\textwidth]{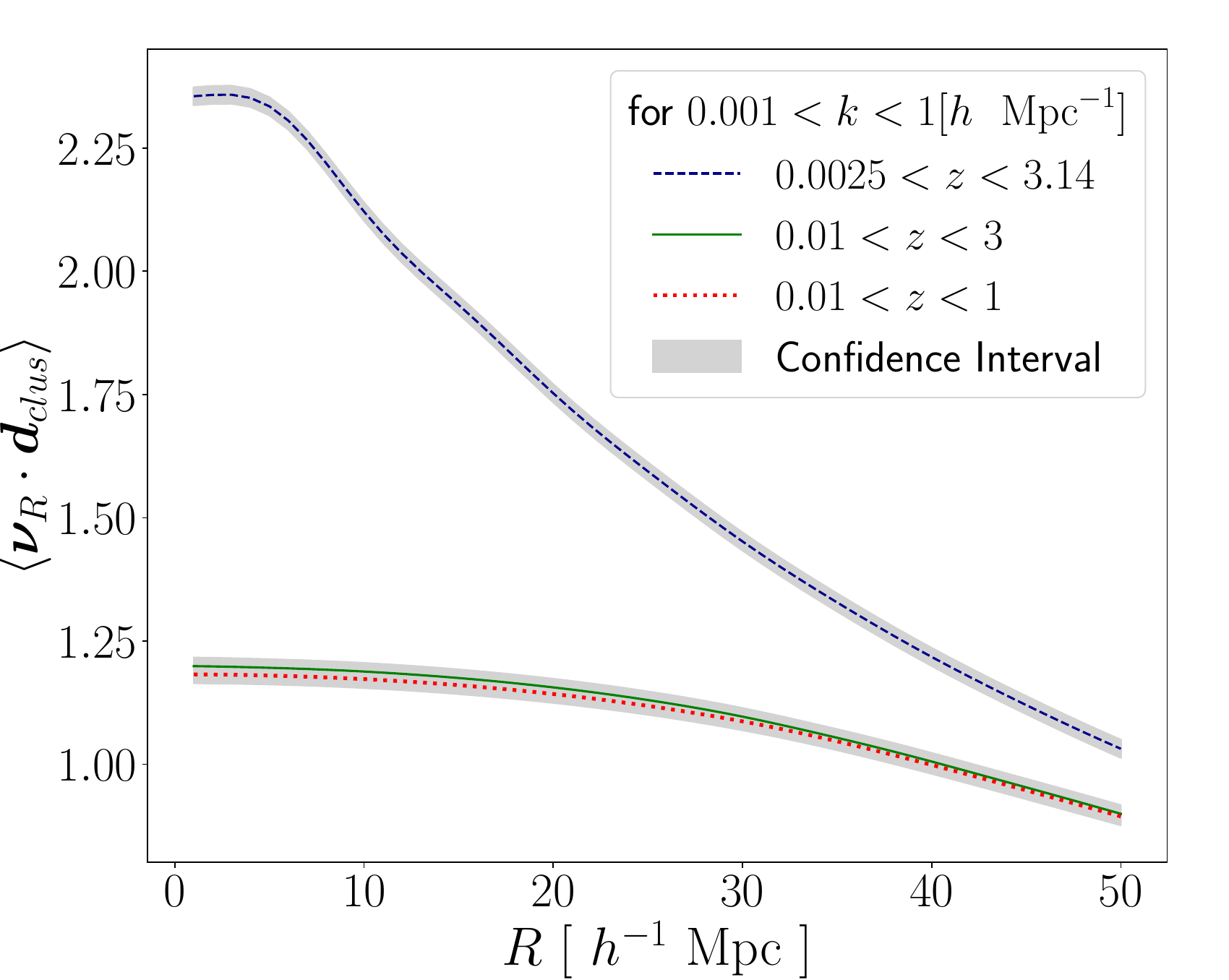}
	\caption{Correlation between the clustering and kinematic dipoles using the One-loop nonlinear power spectrum as a function of bulk radius. This graph is plotted for three redshift ranges: 
		$0.0025 < z < 3.14$ (blue dashed line), 
		$0.01 < z < 3$ (green solid line), 
		and 
		$0.01 < z < 1$ (red dotted line). The gray bars represent the confidence intervals calculated from the error bars.\label{fig:CConeloop50_data,k=1,0.0025,0.01}}
\end{figure}
We have $d_c=d_{\text{kin}} r_c\sigma_S/\sigma_c$ and $\sigma^2_c=A_{\text{kin}}^2 \langle \bf{v} \cdot \bf{v} \rangle/(3c^2)$, where $A_{\text{kin}}=2+\bar{x}(1+\bar{\alpha})$ and $r_c$ is the correlation coefficient between the kinematic and clustering dipoles, defined as:
\begin{equation}
    \label{eq:coefficient}
    r_c=\frac{\langle \mathbf{d}_{\text{kin}} \cdot \mathbf{d}_\text{S} \rangle}{\sqrt{\langle \mathbf{d}_{\text{kin}} \cdot \mathbf{d}_{\text{kin}} \rangle \langle \mathbf{d}_\text{S} \cdot \mathbf{d}_\text{S} \rangle}}.
\end{equation}
Considering that the peculiar velocity, $v_{\text{kin}}$, is proportional to the kinematic dipole and that the shot-noise dipole is independent of the kinematic term, this coefficient can be written as:
\begin{equation}
    \label{eq:c_coefficient}
    r_c=\frac{\langle \mathbf{v}_{\text{kin}} \cdot \mathbf{d}_\text{clus} \rangle}{\sqrt{\langle \mathbf{v}_{\text{kin}} \cdot \mathbf{v}_{\text{kin}} \rangle \langle \mathbf{d}_\text{S} \cdot \mathbf{d}_\text{S} \rangle}}.
\end{equation}
The total peculiar velocity $\textbf{v}_{\text{kin}}$ includes two terms. the first term is a smooth velocity $\textbf{v}_{\text{R}}$ associated with our Local Group and induced by LSSs, which correlates with the clustering dipole. The second term is a stochastic component, $\textbf{v}_{\text{s}}$, which arises from the remaining random motions within the local distribution. These two velocity components are uncorrelated, similar to the components of the source-based dipole, and thus $\langle \mathbf{v}_\text{kin} \cdot \mathbf{v}_\text{kin} \rangle = \langle \mathbf{v}_\text{R} \cdot \mathbf{v}_\text{R} \rangle + \sigma^2_{\text{s}}$, and $\langle \mathbf{d}_\text{S} \cdot \mathbf{d}_\text{S} \rangle = \langle \mathbf{d}_\text{clus} \cdot \mathbf{d}_\text{clus} \rangle + \langle \mathbf{d}_\text{SN} \cdot \mathbf{d}_\text{SN} \rangle$.
The bulk velocity can be computed by solving the Euler equations, following the $\Lambda$CDM model \citep{2014IJMPD..2342025B}:
\begin{equation}
    \label{eq:bulk-velocity}
    \langle \mathbf{v}_R \cdot \mathbf{v}_R \rangle = \frac{H_0^2f^2}{2\pi^2} \int P(k) W_v^2(kR) dk,
\end{equation}
where $W_v(kR)$ is the Fourier transformation of the window function 
and $f$ indicates the matter growth rate: 
\begin{equation}
    f = \frac{d \, \ln \delta_m}{d \, \ln a} \approx \Omega_m^{0.55}, \quad \text{where} \quad \Omega_m = \frac{\rho_m}{\rho_m + \rho_\Lambda}.
\end{equation}
Here, $\rho_m$ and $\rho_\Lambda$ represent the density of dark matter and dark energy.
To determine the correlation coefficient, we require the terms: $\langle \mathbf{v}_{\text{R}} \cdot \mathbf{d}_{\text{clus}} \rangle$, the variance of $\mathbf{v}_{\text{R}}$ and $\mathbf{v}_{\text{s}}$ as well as clustering and shot-noise dipoles. The correlation term is derived using the relations for and the bulk velocity and clustering dipole that we established above and \ref{ssec:clustering}, respectively:
\begin{align}
    \label{eq:v_z.d_z}
    \langle \mathbf{v}_R \cdot \mathbf{d}_{\text{clus}} \rangle &= 3f H_0 \int \frac{dk}{2\pi^2} k W_v(kR) P(k) \notag \\
    & \cdot \int dz \frac{b(z)}{N} \frac{dN}{dz} D(z) j_{1}[k\chi(z)].
\end{align}
This correlation is illustrated in Figure~\ref{fig:CConeloop50_data,k=1,0.0025,0.01} for the $\Lambda$CDM model, using the NVSS fiducial redshift distribution.
Considering a lower redshift interval for the distribution of sources, we observe a significant increase in correlation. This outcome aligns with our expectations, since anisotropy tends to increase in local environments.

\section{Beyond $\Lambda$CDM model}\label{sec:Beyond}
Everything discussed so far has been a reconstruction of $\Lambda$CDM predictions regarding the amplitude of the NVSS dipole, including nonlinear corrections. There is also the consideration that the remaining discrepancy between observed and expected dipoles could be addressed with beyond $\Lambda$CDM physics.
\subsection{Long-Mode Modulation}\label{ssec:L-mode modulation}
{{The mismatch between the CMB and LSS rest frames can indicate the presence of an intrinsic dipole. \cite{1988ASPC....4..344G} reviewed observations that showed a discrepancy between peculiar velocities and the mass distribution inferred from galaxy surveys, and then suggested that a superhorizon curvature perturbation could explain the bulk flow anomaly. This idea was further developed by \cite{1991PhRvD..44.3737T}, who proposed a tilted Universe in which isocurvature perturbations on scales larger than the Hubble radius at the onset of inflation create an intrinsic dipole in the CMB rest frame. \cite{1991PhRvD..44.3737T} explained that if the inflationary epoch lasted about ten e-folds longer than necessary to solve the horizon and flatness problems, such isocurvature superhorizon modes could create a density gradient, making the Universe appear “tilted” in the CMB rest frame.  

The effects of superhorizon perturbations on the CMB have also been examined in \citep{1996PhRvD..53.2908L, 2008PhRvD..78l3529Z, 2008PhRvD..78h3012E, 2008PhRvD..78l3520E, 2009PhRvD..80h3507E}, indicating that adiabatic modes do not create a substantial effect on the CMB dipole. In a more recent study, \cite{2022JCAP...10..019D} quantified the idea of tilted Universe within a perturbed FLRW framework and showed that neither isocurvature nor adiabatic superhorizon fluctuations significantly affect the galaxy number count dipole. Their findings also indicate that an isocurvature superhorizon mode can reduce the CMB dipole by providing a negative contribution to the dipole. While this could explain the mismatch between the CMB and LSS rest frames, it would imply that their alignment is accidental and motivates the search for alternative explanations.

In this work, we employ a phenomenological long-mode modulation to study its impact on matter distribution, without assuming a specific physical origin. Physical mechanisms that could generate such modulations have been discussed in the literature \citep[e.g.,][]{2008PhRvD..78l3520E, 2009PhRvD..80h3507E, 2010PhLB..683..298D, 2013JCAP...08..007L}. These early Universe ideas suggest that superhorizon curvaton perturbation can create the power asymmetry. However, their imprint on LSS is insignificant in perturbed FLRW calculations \citep{2022JCAP...10..019D}. 
As summarized in a recent review by \cite{2025arXiv250523526S}, alternative beyond FLRW models (such as certain Bianchi type extensions, Kantowski–Sachs, and Szekeres models) as well as nonlinear structure growth have also been proposed in the literature. However, as Secrest et al. mentioned, these ideas require further observations and evaluation to form a consistent cosmological model that addresses dipole tension.

In the following, we introduce a tilted-universe scenario in which a long-mode modulation of the primordial curvature perturbation creates the observed power asymmetry in matter distribution. Such adiabatic modulation breaks the exact statistical isotropy, causing departures from the isotropic FLRW paradigm.}}

To build the power asymmetry, one approach is to take inspiration from the method used to explain the hemispherical power asymmetry in CMB anisotropies \citep{Gordon2007, Eriksen2007}. In this method, the simple parameterization of dipolar power asymmetry in temperature anisotropy that is induced by a long-mode modulation is given by:
\begin{equation} 
    \label{eq:T_anis}
    T(\hat{\mathbf{n}}) = \bar{T}(\hat{\mathbf{n}}) [1+ A_d \cos(\theta_{\hat{\mathbf{n}} . \hat{\mathbf{p}}})],
\end{equation}
where $\bar{T}(\hat{\mathbf{n}})$ shows isotropic part of the temperature fluctuation and $A_d$ represents amplitude of long-mode modulation. In Equation~(\ref{eq:T_anis}), $\hat{\mathbf{p}}$ denotes the preferred direction, and $\hat{\mathbf{n}}$ represents the observational direction. In what follows, we adopt this idea, originally proposed in a different context, to address the remaining discrepancy between the measured NVSS and the observed CMB dipole by applying a long-mode modulation to the primordial curvature perturbation. The effect of this long mode on LSS is studied extensively as well \citep{2018JCAP...01..051A}. 

Measurements by \cite{PlanckCollaboration2016} suggest that the significance of the observed hemispherical power asymmetry in the CMB inclines at larger scales ($\ell \lesssim 65$). 
Given this scale dependency, we can model the initial adiabatic curvature perturbations as the sum of a large-scale dipole-modulated component and a small-scale statistically isotropic component \citep{2017PhRvD..95f3011Z}:
\begin{equation} 
    \label{eq:R(x)}
    \Tilde{\mathcal{R}}(\mathbf{x}) = \Tilde{\mathcal{R}}^{lo}(\mathbf{x}) + \mathcal{R}^{hi}(\mathbf{x}),
\end{equation}
the superscript 'lo' refers to large-scale modes (low wave numbers), while the 'hi' indicates small-scale modes. In contrast to \cite{2017PhRvD..95f3011Z}, which reconstructs the physical modulation of primordial fluctuations based on the large-scale CMB temperature dipolar asymmetry, we use this modulation to build a model that assigns the remaining discrepancy to intrinsic anisotropies. Based on this model, we then find an upper limit for the amplitude of this intrinsic dipolar anisotropy in NVSS.

\subsubsection{A Simple Model}\label{sssec:SimpleModel}
Following the introduced mechanism, the curvature perturbation associated with large-scale modes can be linearly modulated as follows:
\begin{equation} 
    \label{eq:modul_R} 
    \Tilde{\mathcal{R}}^{lo}(\mathbf{x}) = \mathcal{R}^{lo}(\mathbf{x}) [1 + A_d \cos(\theta_{\hat{\mathbf{n}}. \hat{\mathbf{p}}})]. 
\end{equation} 
In this modulation, the preferred direction is the same as the CMB's dipole direction; for simplicity, we consider it aligned with the $\hat{z}$ axis. Additionally, in this simple model, we assume that the modulation passes through the center of the Milky Way.
The power spectrum of the modulated curvature perturbation can be derived using Equation~(\ref{eq:modul_R}) as:
\begin{align}
    \label{eq:pp}
    \notag
    \langle \Tilde{\mathcal{R}}(\mathbf{k}) \tilde{\mathcal{R}}^*(\mathbf{k}^{\prime}) \rangle =& \frac{2\pi^2}{k^3} \mathcal{P}^{\Lambda CDM}_{\mathcal{R}}(k) \\
    &\times\delta^3(\mathbf{k}-\mathbf{k}^{\prime})\big[1+2A_d \cos(\theta_{\hat{\mathbf{n}}.\hat{\mathbf{p}}})\big],
\end{align}
where we neglected the term proportional to $A_d^2$. 
By deriving the modified power spectrum from this, we can then calculate the clustering.

To estimate a higher limit for the amplitude of this modulation, we attribute the remaining discrepancy between the observed and modeled NVSS dipoles (Section~\ref{sec:Results}) in the direction of the CMB dipole to an intrinsic anisotropy. We compare the modeled NVSS dipole, whose kinematic, clustering and shot-noise components are introduced in Sections \ref{ssec:Kinematic}, \ref{ssec:clustering} and \ref{ssec:Shot-noise}, with the total NVSS dipole reported by \cite{Secrest2022}: $|d| = (1.23 \pm 0.25) \times 10^{-2}$, about $45^\circ$ away from the CMB dipole direction. 

To find a more accurate upper limit for the modulation amplitude $A_d$, we use the NVSS dipole obtained by the minimum model of \cite{Cheng2024}:
$|d| = (0.61^{+0.26}_{-0.24}) \times 10^{-2}$ with an angular deviation of $28^{+25}_{-18}\,\mathrm{^\circ}$ from the CMB dipole. This choice is motivated by the fact that our clustering values are based on the fiducial model for redshift distribution (Sections \ref{ssec:clustering} and \ref{sec:Results}), so the clustering contribution for radio galaxies is larger than in their minimum model (as explained in Section \ref{ssec:clustering}, the fiducial model is obtained by scaling the minimum model to better match low-redshift tomographic bias estimates). This increases the total dipole amplitude and reduces the residual discrepancy in amplitude, making the minimum model result of \cite{Cheng2024} more suitable for estimating a conservative upper limit on $A_d$. Using the residual amplitude along the CMB dipole direction, we find
$A_d \lesssim 0.22.$ If instead we adopt the fiducial model, with $|d| = 0.67 \times 10^{-2}$ and an angular deviation of about $33^\circ$ from the CMB dipole, the inferred amplitude decreases to $A_d \simeq 0.12$.

This approach cannot be applied to our CatWISE sample of quasars to determine the upper limit for $A_d$, as we used a redshift distribution for this sample that minimized the contributions from local clustering. Consequently, the clustering dipole amplitudes we calculated for the quasars are significantly smaller than the clustering values from the NVSS (Section~\ref{sec:Results}), making them unsuitable for the mechanism described earlier.
\begin{figure}[!htb]
     \centering
  \includegraphics[width=0.45\textwidth]{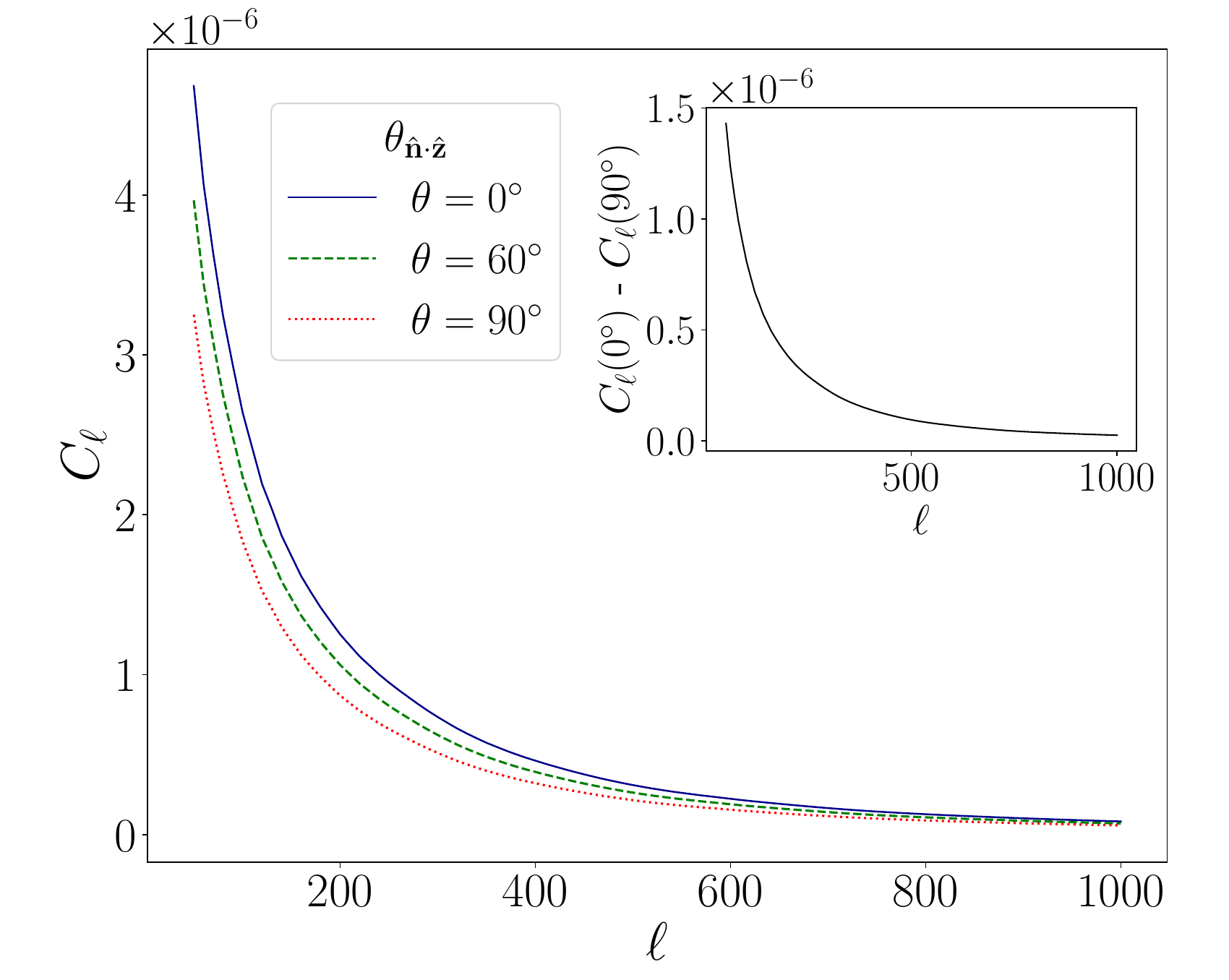}
    \caption{Angular power spectrum of clustering for 
    $\ell$ 
    values greater than 
    $50$. The blue line represents clustering in the direction of modulation, while the green dashed line and red dotted line represent clustering at angles of 
    $60^\circ$ 
    and 
    $90^\circ$ 
    relative to the 
    $\hat{z}$ 
    direction, respectively. These clustering values are calculated using the fiducial NVSS redshift distribution and the redshift interval of $0.01$ to $3.14$.
    }
    \label{fig:c_ell}
\end{figure}
To calculate the clustering dipole (Section~\ref{ssec:clustering}), we need the clustering angular power spectrum at 
$\ell=1$, but since the dipole modulation breaks the isotropy assumption, $C_1$ of the angular power spectrum represents two points in the sky with different power spectra. To resolve this issue, we consider higher $\ell$ values. These $\ell$ values should be sufficiently large such that the angle between the two points for which we are calculating the angular power spectrum is small enough for both to fall within a region (with equal-time and linear power spectrum).
Finally, in Figure~\ref{fig:c_ell}, we illustrate the clustering for a modulation amplitude $A_d \simeq 0.22$, shown for three different directions relative to the CMB dipole direction. The inset plot shows the difference between the clustering in the direction of long-mode modulation and the clustering at a 90-degree angle from it.

\subsection{Modified Gravity}\label{ssec:ModifiedGravity}
\vspace{-4.5em}
\begin{center}
\begin{table*}[t]
\begin{tabular}{|c|c|c|c|c|c|}
    \hline
    {Galaxy Sample} & {Redshift} 
    & \multicolumn{2}{c|}{$0.01 < z < 3$} 
    & \multicolumn{2}{c|}{$0.0025 < z < 3.14$} \\
    \hline
    \multirow{2}{*}{\shortstack{NVSS\\ \scriptsize(Radio Galaxies)}} 
    & $P(k)$ & Linear & Nonlinear & Linear & Nonlinear \\
    \cline{2-6}
    & $d_{\text{clus}}(\times 10^{-2})$ 
    & $0.44 \pm 0.01$ & $0.44 \pm 0.02$ 
    & $0.67 \pm 0.03$ & $0.69 \pm 0.05$ \\

    \hline
    \multirow{3}{*}{\shortstack{CatWISE\\ \scriptsize(Quasars)}} 
      & {Redshift} & \multicolumn{4}{c|}{$0 \lesssim z < 3$} \\
    \cline{2-6}
      & $P(k)$ & \multicolumn{2}{c|}{Linear} & \multicolumn{2}{c|}{Nonlinear} \\
    \cline{2-6}
      & $d_{\text{clus}}(\times 10^{-2})$ 
      & \multicolumn{2}{c|}{$0.030 \pm 0.004$} 
      & \multicolumn{2}{c|}{$0.032 \pm 0.005$} \\
     
    \hline
\end{tabular}
\caption{Clustering dipole values from the angular power spectrum (\(f(R)\) Hu-Sawicki model, \(|f_{R0}|=5\times10^{-5}\))}
\label{tab:clustering_dipole_CCL_MG1}
\end{table*}
\end{center}
\begin{center}
\begin{table*}[t]
\begin{tabular}{|c|c|c|c|c|c|}
    \hline
    {Galaxy Sample} & {Redshift} 
    & \multicolumn{2}{c|}{$0.01 < z < 3$} 
    & \multicolumn{2}{c|}{$0.0025 < z < 3.14$} \\
    \hline
    \multirow{2}{*}{\shortstack{NVSS\\ \scriptsize(Radio Galaxies)}} 
    & $P(k)$ & Linear & Nonlinear & Linear & Nonlinear \\
    \cline{2-6}
    & $d_{\text{clus}}(\times 10^{-2})$ 
    & $0.44 \pm 0.01$ & $0.44 \pm 0.01$ 
    & $0.65 \pm 0.04$ & $0.66 \pm 0.04$ \\

    \hline
    \multirow{3}{*}{\shortstack{CatWISE\\ \scriptsize(Quasars)}} 
      & {Redshift} & \multicolumn{4}{c|}{$0 \lesssim z < 3$} \\
    \cline{2-6}
      & $P(k)$ & \multicolumn{2}{c|}{Linear} & \multicolumn{2}{c|}{Nonlinear} \\
    \cline{2-6}
      & $d_{\text{clus}}(\times 10^{-2})$ 
      & \multicolumn{2}{c|}{$0.030 \pm 0.003$} 
      & \multicolumn{2}{c|}{$0.031 \pm 0.003$} \\
     
    \hline
\end{tabular}
\caption{Clustering dipole values from the angular power spectrum (\(f(R)\) Hu-Sawicki model, \(|f_{R0}|=5\times10^{-6}\))}
\label{tab:clustering_dipole_CCL_MG2}
\end{table*}
\end{center}
\hspace*{\parindent}To extend our analysis of the clustering dipole beyond the predictions of the standard cosmological model, we examine modified gravity theories that could influence structure formation on cosmic scales. Specifically, we focus on the \( f(R) \) gravity model, using the formulation by \cite{Hu2007}, which modifies the functional form of the Ricci scalar \( R \) to allow deviations from general relativity. This approach enables us to study potential changes in the clustering dipole across both linear and nonlinear regimes, providing insights into how alternative gravity models might impact large-scale structures and offering a critical test of the standard cosmological framework.
The Hu-Sawicki model has been widely studied and shown to be compatible with a range of observational data, including solar system tests, large-scale structure, cosmic microwave background measurements, and galaxy cluster dynamics \citep{Hu2007, Pogosian08, Lombriser2012}. These works demonstrate that the model can successfully mimic the expansion history of \(\Lambda\)CDM while allowing for distinguishable deviations in the growth of structure, making it a well-motivated candidate for probing modifications to gravity.

\cite{Hu2007} modified the Einstein-Hilbert action as follow:
\begin{equation}
    \label{eq:action}
    S=\int \text{d}^4x\sqrt{-g}\left [\frac{R+f(R)}{2\kappa^2}+\mathcal{L}_m\right ],
\end{equation}
where \(R\) is Ricci scalar, \(\kappa^2\equiv8\pi G\) and \(\mathcal{L}_m\) represents matter Lagrangian.
The functional form of \(f(R)\) was chosen to satisfy specific observational criteria: at high redshifts, it should reproduce \(\Lambda\)CDM behavior, consistent with constraints from the CMB. At low redshifts, it drives accelerated expansion in a manner close to \(\Lambda\)CDM, but without a true cosmological constant. The parametrization of \( f(R) \) provides sufficient degrees of freedom to capture a range of low-redshift phenomena, while retaining \(\Lambda\)CDM behavior as a limiting case to enable cosmological and solar-system tests of deviations from general relativity.
To meet these requirements, \( f(R) \) is constructed to satisfy the following asymptotic conditions:
\begin{equation}
    \label{eq:limit}
    \lim_{R \to \infty} f(R) = \text{const.}, \quad \lim_{R \to 0} f(R) = 0.
\end{equation}
These conditions can be achieved by employing a general class of broken power-law models for \( f(R) \).
\begin{equation}
    \label{eq:f(R)}
    f(R)=-m^2\frac{c_1(R/m^2)^n}{c_2(R/m^2)^n+1},
\end{equation}
where \(c_1\), \(c_2\), and \(n\) are dimensionaless parameters, and \(m\) is a curvature scale defined as
\begin{equation}
    \label{eq:m}
    m^2 \equiv \frac{\Omega_m H_0^2}{c^2},
\end{equation}
here, \(\Omega_m\) represents the current fractional density, and \(H_0\) is the Hubble constant.

To investigate the clustering dipole in this model, both the linear and nonlinear matter power spectra are required. 
Our analysis is built upon the background cosmological parameters adopted from the \cite{Euclid2024}: $\{\Omega_m, \Omega_b, h, n_s, \sigma_8\} = \{0.33, 0.05, 0.67, 0.96, 0.853\}$. In the Hu-Sawicki \( f(R) \) gravity model, the parameter \(f_{R0}\) characterizes the present-day value of the scalar degree of freedom and determines the strength of the deviation from general relativity at late times; more negative values correspond to stronger modifications of gravity. \\ 
The choice of $f_{R0}$ is subject to a hierarchy of observational constraints across different scales. The tightest bounds originate from galactic-scale tests: analyses of stellar-gas offsets and galactic rotation curves require $|f_{R0}| \lesssim 10^{-8}$ \citep{Desmond2020, Landim2024}. On cosmological scales, cluster abundance measurements provide progressively weaker but still significant limits: combining cluster counts with weak lensing mass calibration yields $|f_{R0}| < 4.8 \times 10^{-6}$ (95\% CL) \citep{SPTcluster2024}, while using cluster counts alone relaxes the constraint to $|f_{R0}| \lesssim 4.9 \times 10^{-5}$ \citep{eRASS1cluster2024}.\\ Given this context, our analysis serves as a proof of concept to explore the sensitivity of the clustering dipole to modified gravity effects. We therefore compute the dipole for two benchmark values: $f_{R0} = -5 \times 10^{-6}$, which probes the regime accessible to cosmological surveys but is excluded by galactic tests, and $f_{R0} = -5 \times 10^{-5}$, which is ruled out by all current data but serves to illustrate the maximum potential effect on our observable. This approach allows us to bracket the possible phenomenological impact and calibrate the response of the clustering dipole across different scales of modified gravity strength. The model exponent is fixed to $n=1$ for this study. \\
With this parameter space defined, we computed the corresponding matter power spectra. The linear power spectrum was obtained using the MGCAMB\footnote{https://github.com/sfu-cosmo/MGCAMB\_v4} code  \citep{Zhao2009, Hojjati2011, Zucca2019, Wang2023}.
To compute the nonlinear matter power spectrum, we used the FREmu\footnote{https://github.com/AstroBai/FREmus} code, which provides an emulator for the Hu–Sawicki \(f(R)\) model. FREmu is calibrated against full N-body simulations, enabling reliable predictions in the nonlinear regime of modified gravity models. The \(k\)-range in this code spans from \(10^{-4}\) to \(0.5 ~h\rm{Mpc}^{-1}\), expressed in units scaled by the Hubble parameter \(h\) \citep{Bai2024, Bai2024*}.

\begin{figure}[ht!]
\includegraphics[width=0.45\textwidth]{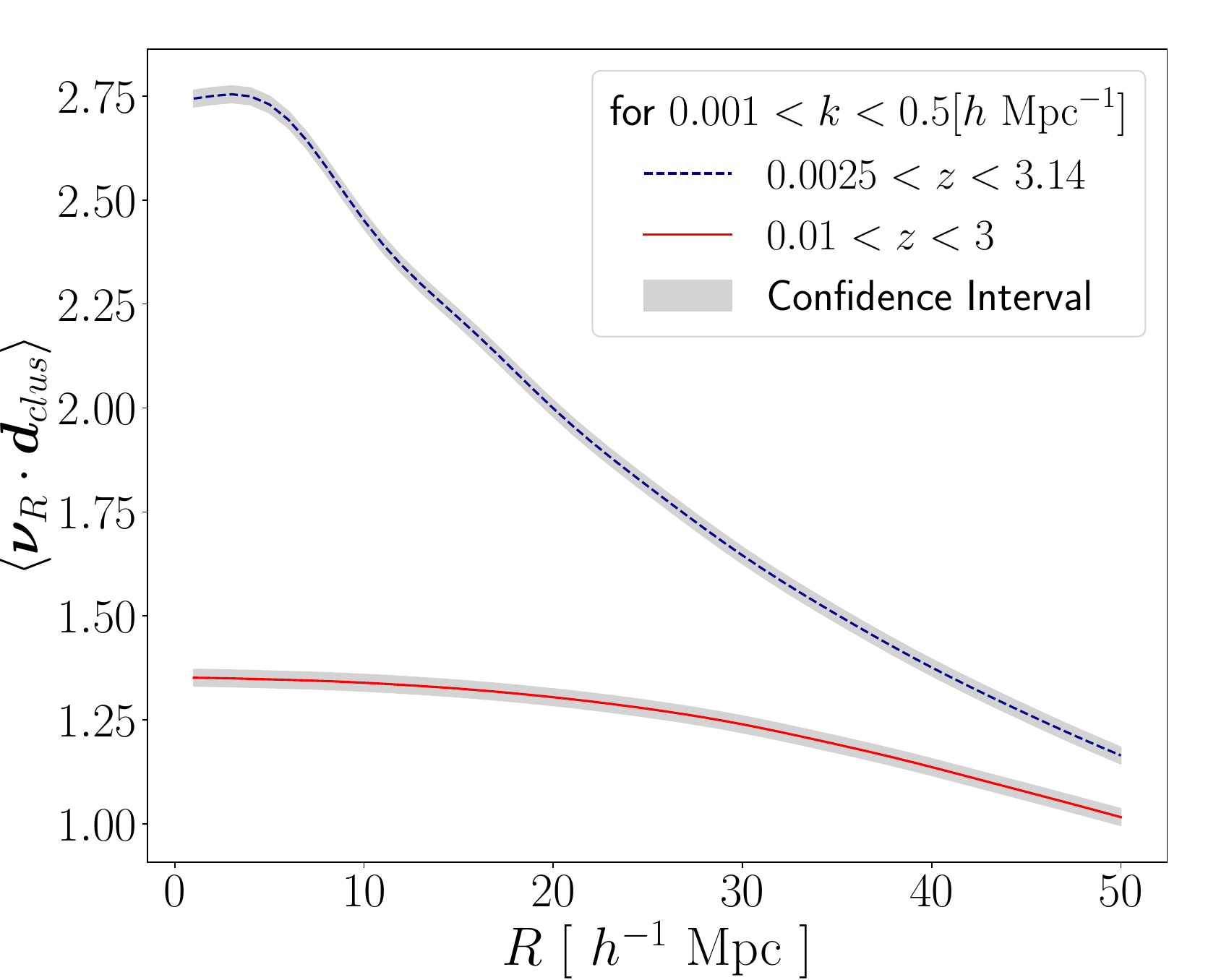}
\caption{Correlation between the clustering and kinematic dipoles using the Hu-Sawicki model as a function of bulk radius in the nonlinear regime. This graph is plotted for two redshift ranges: 
    $0.0025 < z < 3.14$ (blue dashed line),  
    and 
    $0.01 < z < 3$ (red solid line). The gray bars represent the confidence intervals calculated from the error bars.\label{fig:CCMG}}
\end{figure}

Having obtained the linear and nonlinear matter power spectra of the Hu–Sawicki \(f(R)\) model, we computed the clustering dipole using the relation given in \ref{ssec:clustering}, which accounts for the contributions from the density fluctuations across different scales. The resulting clustering dipole amplitudes for the ${f_{R0} = -5 \times 10^{-5}}$ case are summarized in Table \ref{tab:clustering_dipole_CCL_MG1} for both the NVSS and CatWISE catalogs. Similarly, the results for the ${f_{R0} = -5 \times 10^{-6}}$ scenario are presented in Table \ref{tab:clustering_dipole_CCL_MG2}. \\ A comparison of the dipole amplitudes in Tables \ref{tab:clustering_dipole_CCL_MG1} and \ref{tab:clustering_dipole_CCL_MG2} reveals a clear trend: the clustering dipole is systematically enhanced in the $f(R)$ gravity scenarios compared to the $\Lambda$CDM. As expected theoretically, this enhancement is more pronounced for the stronger modification ($|f_{R0}| = 5 \times 10^{-5}$) than for the weaker one ($|f_{R0}| = 5 \times 10^{-6}$). This behavior aligns with the physical expectation that a larger $|f_{R0}|$ leads to stronger fifth forces and more efficient growth of structure, thereby increasing the amplitude of the clustering dipole. 

To better compare the standard model and the Hu-Sawicki modified gravity model, we calculated Equation~(\ref{eq:v_z.d_z}), which is shown in Figure~\ref{fig:CCMG}, illustrating the correlation between the kinematic dipole and the clustering dipole derived from the NVSS catalog in the nonlinear regime.

\section{Conclusion \& Discussion} \label{sec:discussion} 
In this work, we revisited the tension between the dipole observed in the CMB and that inferred from the LSS of the Universe, focusing on the NVSS radio galaxy catalog, with the CatWISE quasar sample also considered. The CMB dipole is interpreted as a kinematic effect arising from our peculiar motion, and a corresponding kinematic dipole is expected to appear in LSS. However, multiple studies have reported an excess dipole amplitude and a directional mismatch in LSS surveys compared to the CMB prediction.

Our analysis incorporates the three key components of the observed dipole—the kinematic, clustering, and shot-noise contributions. We extended previous linear analyses by including nonlinear corrections using both standard perturbation theory and the halo model. These refinements allow for a more accurate treatment of the clustering dipole and its correlation with the kinematic term. We find that nonlinear effects moderately enhance the clustering dipole amplitude but are insufficient on their own to resolve the observed discrepancy.

To compute clustering dipole amplitude, we utilized the data-driven approach developed by \cite{Cheng2024}, which directly infers the bias function $ b(z) $ and the differential source counts $dN/dz$ for the NVSS sample using the Tomographer and cross-matching techniques. This approach differs from previous studies that employed separate parametric functional forms for $b(z)$ and $dN/dz$. 
The amplitude of the NVSS dipole reported by \cite{Cheng2024} for their fiducial model is $|d|= (0.67^{+0.31}_{-0.28}) \times 10^{-2}$. This result deviates from the measured dipole by \cite{Secrest2022} by about $1.41 \sigma$, where both studies utilized the joint probability distribution function of dipole amplitude and angular distance from the CMB dipole. 

\cite{Cheng2024} estimated the amplitude of the NVSS clustering dipole to be $|d|= (0.43 \pm 0.07)  \times 10^{-2}$ for their fiducial model. In our analysis, we considered the contribution of low-redshift sources ($z<0.01$) in the fiducial redshift distribution, which enhances the clustering dipole amplitude. Utilizing the one-loop model, our clustering dipole is at least $12\text{-}16\%$ (linear/nonlinear) higher than \cite{Cheng2024}, while with Halofit, it is at least $20\text{-}22\%$ higher than their reported value in the fiducial redshift distribution (see Tables \ref{tab:clustering_dipole_Approx_unequaltime_one_loop} and \ref{tab:clustering_dipole_CCL}). This clustering enhancement increases the modeled NVSS total dipole amplitude, reducing the deviation from the observed dipole relative to the $1.41\sigma$ difference reported above.

We also repeated our clustering analysis for the CatWISE quasar selection (Section \ref{sec:Results}). Our clustering values are consistent with previous studies \citep{2021ApJ...908L..51S}. As reported in Tables \ref{tab:clustering_dipole_Approx_unequaltime_one_loop} - \ref{tab:clustering_dipole_CCL_MG2}, the clustering dipole amplitude for this sample is noticeably lower than that of NVSS.  This is expected because the sample contains very few nearby sources, reducing the impact of local clustering on the measured dipole. For this reason, the clustering and kinetic correlation are negligible in this case. However, as \cite{Cheng2024} notes, the source number densities in the CatWISE catalog are comparable to those in NVSS, which results in similar shot-noise levels and therefore affects the dipole estimates from this data set.

We also examined the correlation between the clustering and kinematic dipoles using NVSS redshift distribution, showing that local structure can induce a positive correlation that increases the expected dipole amplitude. This effect reduces part of the tension, particularly at low redshifts where anisotropies are stronger. \\
Beyond $\Lambda$CDM, we considered two complementary scenarios one with an origin in the early Universe and the other in the late-time.  {{In the early Universe case}}, we studied intrinsic anisotropy through long-mode modulation of primordial perturbations, motivated by the hemispherical power asymmetry in the CMB. Using the residual discrepancy between observed and predicted dipoles, we derived an upper limit of $A_d \lesssim0.22$ ~for the amplitude of such intrinsic anisotropy in the NVSS dipole.

{{For the late-time scenario, we investigated modified gravity within the Hu–Sawicki $f(R)$ model, in which the enhanced growth of structure leads to a larger clustering dipole. Using MGCAMB for the linear regime and FREmu for nonlinear corrections, our calculations show that for $|f_{R0}| = 5 \times 10^{-5}$ the enhancement is $\sim 28\%$ for both regimes, while for $|f_{R0}| = 5 \times 10^{-6}$ the enhancement is $\sim 22\%$ (linear) and $\sim 24\%$ (nonlinear) (see Tables \ref{tab:clustering_dipole_CCL_MG1} and \ref{tab:clustering_dipole_CCL_MG2}). These enhancements reduce the deviation from the measured NVSS dipole amplitude reported by \cite{Secrest2022}, suggesting that inclusion of nearby sources has a non-negligible effect in reconciling the inconsistency between the model and data.}}

In summary, our results indicate that while nonlinear clustering and correlations with local bulk flows alleviate the dipole tension within $\Lambda$CDM, they do not fully resolve it. Models with intrinsic anisotropies or modified gravity, such as $f(R)$, can further reduce the discrepancy and offer testable predictions for future surveys. Upcoming wide-area radio surveys (e.g., SKA) and deep spectroscopic programs will be critical for refining clustering measurements and distinguishing between $\Lambda$CDM and beyond-$\Lambda$CDM explanations of the dipole tension.
	\section*{ACKNOWLEDGMENTS}
We are grateful to the anonymous referee for careful and insightful comments that improved both the presentation and the results of this paper. \\
We thank Adam Lidz and Yun-Ting Cheng for providing the NVSS processed data. 
We thank Elisabeth Krause for advice regarding the use of the Core Cosmology Library (CCLX) package at the ICTP 2024 Cosmology Summer School and Dominik Schwarz for the helpful comment at the EAS 2024 Conference. 
We also thank Jiachen Bai and collaborators for helpful clarifications regarding the use of the FREmus code, and Sebastian von Hausegger for helpful comments on the earlier version of this manuscript. \\
SB is partially supported by the Abdus Salam International Center for Theoretical Physics (ICTP) under the regular associateship scheme. Moreover, we are all partially supported by the Sharif University of Technology Office of Vice President for Research under Grant No. G4010204.


\bibliography{Anisotropy}{}
\bibliographystyle{aasjournal}
\appendix
\section{Nonlinear Regime}\label{sec:Nonlinear}
As mentioned in \ref{ssec:clustering}, calculating the clustering dipole requires the power spectrum in both linear and nonlinear regimes. For the nonlinear regime, we employ two approaches: Standard Perturbation Theory (SPT) and the Halo Model. In the following, we briefly introduce these frameworks that are used to calculate the clustering dipole and its correlation to velocity in Section~\ref{sec:Results}.
\subsection{Standard Perturbation Theory}
According to SPT, the matter power spectrum can be expressed as a sum of loop corrections. Specifically, it is written as \(P(k) = P^{(0)}(k) + P^{(1)}(k) + \ldots, \) where \( P^{(0)}(k) = P_{\text{lin}}(k) \) represents the linear term, and \( P^{(1)}(k) \) is the 1-loop correction. These terms capture the gravitational interactions beyond the linear regime and are necessary to account for the nonlinear growth of structure \citep{Bernardeau2002}. The 1-loop correction is composed of two primary terms, \( P_{13}(k) \) and \( P_{22}(k) \), defined as \(\ P^{(1)}(k) = 2P_{13}(k) + P_{22}(k).\)
In this expression, \( P_{13}(k) \) corresponds to the correction of the linear propagator, which arises from the linear evolution of modes, while \( P_{22}(k) \) represents mode coupling between different Fourier modes \citep{Fry1994}.
The term \(P_{13}(k)\) is given by:
\begin{equation}
\label{eq:P13}
P_{13}(k) = 3P_\text{lin}(k) \int d^3\mathbf{q} \, F_3(\mathbf{k}, \mathbf{q}, -\mathbf{q}) P_\text{lin}(q),
\end{equation}
where \(F_3(\mathbf{k}, \mathbf{q}, -\mathbf{q})\) is the kernel that describes the coupling between modes with wave vectors \(\mathbf{k}\) and \(\mathbf{q}\). This kernel captures the three-point correlations and is essential for understanding how linear modes interact.
The term \(P_{22}(k)\) is defined as:
\begin{equation}
P_{22}(k) = 2\int d^3\mathbf{q} \, F_2^2(\mathbf{q}, \mathbf{k-q}) P_\text{lin}(q) P_\text{lin}(|\mathbf{k-q}|),
\label{eq:P22}
\end{equation}
where \(F_2(\mathbf{q}, \mathbf{k-q})\) describes the coupling between two different modes, \(\mathbf{q}\) and \(\mathbf{k-q}\). This term is crucial for capturing the nonlinear interactions between pairs of modes.
The kernels of Equations (\ref{eq:P13}) and (\ref{eq:P22}) are expressed as,
\begin{align}
    \label{eq:Fn}
    F_n(\mathbf{q}_1, \ldots, \mathbf{q}_n) = & \sum_{m=1}^{n-1} \frac{G_m(\mathbf{q}_1, \ldots, \mathbf{q}_m)}{(2n+3)(n-1)}\\ 
    & \times [(2n+1) \alpha(\mathbf{k}, \mathbf{k}_1) F_{n-m}(\mathbf{q}_{m+1}, \ldots, \mathbf{q}_n) \nonumber \\
   & + 2 \beta(\mathbf{k}, \mathbf{k}_1, \mathbf{k}_2) G_{n-m}(\mathbf{q}_{m+1}, \ldots, \mathbf{q}_n) \nonumber ],
\end{align}
and
\begin{align}
    \label{eq:Gn}
    G_n(\mathbf{q}_1, \ldots, \mathbf{q}_n) = & \sum_{m=1}^{n-1} \frac{G_m(\mathbf{q}_1, \ldots, \mathbf{q}_m)}{(2n+3)(n-1)}\\
    & \times [ 3 \alpha(\mathbf{k}, \mathbf{k}_1) F_{n-m}(\mathbf{q}_{m+1}, \ldots, \mathbf{q}_n) \nonumber \\
    & + 2n \beta(\mathbf{k}, \mathbf{k}_1, \mathbf{k}_2) G_{n-m}(\mathbf{q}_{m+1}, \ldots, \mathbf{q}_n) \nonumber ],
\end{align}
with \(F_1=1\) and  \(G_1=1\). Also, the functions \(\alpha\) and \(\beta\) are defined as,
\begin{gather}
    \alpha(\mathbf{k},\mathbf{k}_1) \equiv \frac{\mathbf{k}\cdot \mathbf{k}_1}{k_1^2}, \label{eq:alpha} \\
    \beta(\mathbf{k},\mathbf{k}_1,\mathbf{k}_2) \equiv \frac{k^2(\mathbf{k}_1\cdot \mathbf{k}_2)}{2k_1^2k_2^2}. \label{eq:Beta}
\end{gather}
The significance of these terms lies in their ability to account for the gravitational interactions that lead to the nonlinear growth of structure. The 1-loop correction allows for a better approximation of the power spectrum, particularly at small scales where these nonlinear effects become prominent. However, SPT also faces challenges at very small scales due to increasing cancellations among loop contributions, making resummation techniques necessary for improved accuracy \citep{Crocce2006}.
The 1-loop corrections in this work are computed using the SPT linear power spectrum that we derived from the transfer function of \cite{Eisenstein1998}, which normalizes the amplitude of perturbations at the horizon scale today.
\subsection{Halo Model}
In this work, we also utilize the halo model to describe the nonlinear evolution of the matter density field. The halo model assumes that every dark matter particle resides within a dark matter halo, and these halos act as the building blocks for structure formation \citep{Cooray2002}. By combining three key components—the halo density profile, the halo mass function, and the halo clustering properties—the model captures both the internal structure of individual halos (the "one-halo" term) and the correlations between different halos (the "two-halo" term). This provides a framework for modeling the nonlinear power spectrum of matter, especially on small scales where nonlinear effects dominate. 
The dark matter power spectrum can be written as \citep{Cooray2002}:
\begin{equation}
    \label{eq:p_dm}
    P(k) = P^{1h}(k) + P^{2h}(k),
\end{equation}
The first and second terms describe the scenarios where the two contributions to the density come from the same halo and from different halos, respectively. These terms are given by
\begin{equation}
    \label{eq:p_halo}
    \begin{aligned}
    P^{1h}(k) &= \int dm \, n(m) \left( \frac{m}{\bar{\rho}} \right)^2 |u(k|m)|^2, \\
    P^{2h}(k) &= \int dm_1 \, n(m_1) \left( \frac{m_1}{\bar{\rho}} \right) u(k|m_1) \\
    & \quad \int dm_2 \, n(m_2) \left( \frac{m_2}{\bar{\rho}} \right) u(k|m_2) P_{hh}(k|m_1, m_2).
\end{aligned}
\end{equation}
where $n(m)$ represents the number density of halos with mass $m$, and $\bar{\rho}$ is the background mean density. The function $u(k|m)$ indicates the Fourier transform of the dark matter distribution within a halo of mass $m$ and is the Fourier transform of a spherically symmetric mass distribution $\rho(r)$. In the relation above, $P_{hh}(k|m_1, m_2)$ corresponds to the power spectrum of halos with masses of $m_1$ and $m_2$. The Halo model aims to describe the nonlinear regime beyond the limitation of perturbation theory. In this regime, the growth of structure is primarily governed by the collapse of regions into halos of dark matter, which cannot be described using linear perturbation theory.
\end{document}